\documentclass[trackchanges]{aastex631}
\usepackage{amsmath}

\begin{document}

\title{KM UMa: An active short-period detached eclipsing binary in a hierarchical quadruple system}

\correspondingauthor{Liying Zhu}
\email{zhuly@ynao.ac.cn}

\author[0000-0002-8320-8469]{Fangbin Meng}
\affiliation{Yunnan Observatories, Chinese Academy of Sciences (CAS), 650216, Kunming, China}
\affiliation{University of Chinese Academy of Sciences, No.1 Yanqihu East Rd, Huairou District, Beijing, PR China, 101408}

\author[0000-0002-0796-7009]{Liying Zhu}
\affiliation{Yunnan Observatories, Chinese Academy of Sciences (CAS), 650216, Kunming, China}
\affiliation{University of Chinese Academy of Sciences, No.1 Yanqihu East Rd, Huairou District, Beijing, PR China, 101408}

\author{Nianping Liu}
\affiliation{Yunnan Observatories, Chinese Academy of Sciences (CAS), 650216, Kunming, China}

\author{Ping Li}
\affiliation{Yunnan Observatories, Chinese Academy of Sciences (CAS), 650216, Kunming, China}
\affiliation{University of Chinese Academy of Sciences, No.1 Yanqihu East Rd, Huairou District, Beijing, PR China, 101408}

\author{Jia Zhang}
\affiliation{Yunnan Observatories, Chinese Academy of Sciences (CAS), 650216, Kunming, China}

\author{Linjia Li}
\affiliation{Yunnan Observatories, Chinese Academy of Sciences (CAS), 650216, Kunming, China}

\author{Azizbek Matekov}
\affiliation{Yunnan Observatories, Chinese Academy of Sciences (CAS), 650216, Kunming, China}
\affiliation{University of Chinese Academy of Sciences, No.1 Yanqihu East Rd, Huairou District, Beijing, PR China, 101408}
\affiliation{Ulugh Beg Astronomical Institute,
Uzbekistan Academy of Sciences, 33  Astronomicheskaya str., Tashkent, 100052, Uzbekistan}



\begin{abstract}
The first detailed photometric and spectroscopic analysis of the G-type eclipsing binary KM UMa is presented, which indicates that the system is a short-period detached eclipsing binary. The radial velocity curves were calculated using the cross-correlation function method based on Large Sky Area Multi-Object Fiber Spectroscopic Telescope, Sloan Digital Sky Survey, and our observations, which determined the mass ratio as $q=0.45\ (\pm0.04)$. Based on the light curves from the Transiting Exoplanet Survey Satellite, other survey data, and our multiband observations, the positive and negative O’Connell effects have been detected evolving gradually and alternately over the last 20 yr, which can be explained by the presence of spots on the primary component. A superflare event was detected in the SuperWASP data on 2007 February 28, further indicating that KM UMa is a very active system. We calculated its energy to be $5\times10^{34}$ erg by assuming it occurred on the primary star. Utilizing hundreds of medium-resolution spectra and one low-resolution spectrum, the equivalent width variations of the $H_{\alpha}$ line were calculated, indicating the presence of a 5.21 ($\pm0.67$) yr magnetic activity cycle. The orbital period variations were analyzed using the O-C method, detecting a long-term decrease superimposed with a periodic variation. The amplitude of the cyclic variation is $0.01124\ (\pm0.00004)$ day, with a period of $33.66\ (\pm 0.0012)$ yr, which exceeds the 5.21 yr activity cycle, suggesting that this is more likely attributable to the light travel time effect of a third body. Simultaneously, a visual companion has been detected based on the Gaia astrometric data, indicating that KM UMa is actually in a 2+1+1 hierarchical quadruple system.

\end{abstract}

\keywords{binaries: close — binaries: eclipsing — stars: evolution - stars: individual (KM UMa)}


\section{Introduction} \label{section1}
The multiplicity of stars is a common phenomenon in the process of stellar formation \citep{2013ARA&A..51..269D}. About $12\%$ of solar-type stars are found in triple or multiple systems \citep{2010ApJS..190....1R, 2014AJ....147...87T}. The frequency of companion stars also increases with stellar mass \citep{2019BAAS...51c.483S}. Multiple systems often have a close binary that forms a hierarchical configuration with additional bodies. This hierarchical structure may facilitate the formation of close binaries through Kozai cycles and tidal friction \citep{2006Ap&SS.304...75E}, and potentially lead to mergers that produce exotic objects such as blue stragglers \citep{2009ApJ...697.1048P}. Triple systems have only one hierarchical configuration (2+1), quadruple systems have two configurations (2+2 and 2+1+1), and higher-order systems have more complex configurations. Multiple  systems are formed through a combination of various fundamental mechanisms, including disk instability, core fragmentation, and \textit{N}-body interactions, to accommodate different hierarchical configurations \citep{2021Univ....7..352T}. The detailed study of more multiple systems is crucial to exploring the formation of binaries and multiple systems.

Detached eclipsing binaries are reliable objects for obtaining stellar physical parameters and testing stellar evolutionary models \citep{2017A&A...608A..62H}. They usually exhibit light variations of the EA or EB type \citep{2009ebs..book.....K}. Many peculiar objects are assumed to evolve from detached binaries, such as near contact binaries, W UMa binaries, and blue stragglers, etc \citep{1989AJ.....97..431E, 1994ASPC...56..228B,2012AJ....144...37Z,2017RAA....17...87Q}. When two components of the system are late-type stars, they always display magnetic activities. In many cases, dark or hot spots cause a distorted light curve, the so-called O'Connell effect \citep{1951PRCO....2...85O}. Due to the rotational modulation of the starspots, in many detached binaries (e.g., RS CVn-type stars), the light curves display periodic variations outside of eclipses \citep{1952ApJ...115..301K,2022MNRAS.512.4835M}. The flare events, chromospheric activity variations, and magnetic activity cycles, etc., are well reported in many systems \citep{2007ApJ...659L.157B,2018MNRAS.474..326D,2019MNRAS.490.5832M}. In this paper, we present the first detailed analysis of KM UMa based on photometric and spectroscopic observations, investigating its structural evolution, period variations, activity, and multiplicity.

KM UMa was discovered to be variable by \citet{1958JO.....41...74W}. \citet{1999IBVS.4810....1E} classified it
to be an EB-type binary with a period of 0.351862 day. 
KM UMa is listed as an X-ray source in the Hamburg/RASS Catalogue \citep{2003A&A...406..535Z}, implying that it is an active system. Several survey telescopes, such as the Large Sky Area Multi-Object Fiber Spectroscopic Telescope \citep[LAMOST;][]{2012RAA....12.1197C}, Sloan Digital Sky Survey \citep[SDSS;][]{2011AJ....142...72E}, Transiting Exoplanet Survey Satellite \citep[TESS;][]{2015JATIS...1a4003R}, and Super Wide Angle Search for Planets \citep[SuperWASP;][]{2006PASP..118.1407P}, make photometric or spectroscopic observations of it. The spectral type of LAMOST is G9, and the atmospheric parameters are $T_{\rm eff}=5353\ (\pm 16)$, $[Fe/H]=0.09\ (\pm 0.009)$, $\log(g)=4.32\ (\pm 0.02)$.

\section{Observations} \label{section2}
\subsection{Photometry}

\begin{figure}[ht!]
\plotone{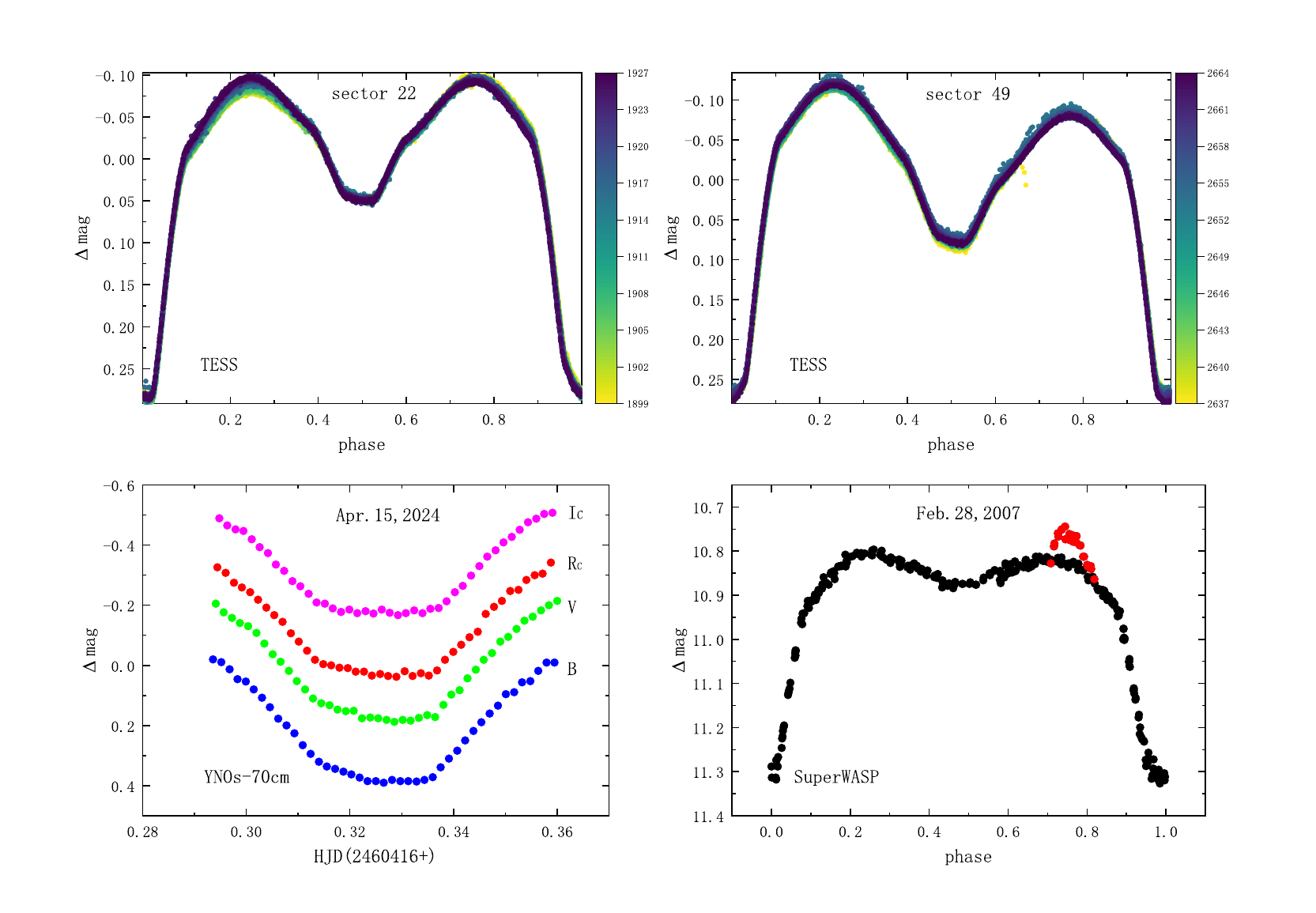}
\caption{Upper two panels: TESS light curves of two sectors. Lower-left panel: multiband eclipse light curves. Lower-right panel: flare light curve discovered by SuperWASP.
\label{fig1}}
\end{figure}

CCD observations of KM UMa have been carried out since 2006 using the 60 cm and 1 m Cassegrain reflecting telescopes at Yunnan Observatories (YNOs) of the Chinese Academy of Sciences. The telescopes are equipped with Andor DW436 2K CCD cameras and standard Johnson-Cousin-Bessel $BVR_{c}I_{c}$ filters. In addition, $BVR_cI_c$ multiband observations of KM UMa were carried out on two nights, 2024 April 15 and 18, using the 70 cm Sino-Thai telescope. This telescope is located at the Lijiang Gaomeigu station of YNOs with an Andor DW936 2K CCD camera. Standard reduction of CCD images was performed using the DAOPHOT package of IRAF software, including bias and flat corrections and aperture photometry. From the observations, 24 primary light minima times and 17 secondary light minima times were obtained by parabolic fitting of the eclipse light variations. The bottom-left panel of Figure \ref{fig1} displays an eclipse light curve in different bands. The photometric data from the YNOs are provided in Table \ref{table_lc}

TESS is an all-sky survey designed to search for transiting exoplanets. It was launched in 2018 April and completes an all-sky survey every two years. TESS divides the entire sky into 26 sectors, each covering 24$^{\circ}$ by 90$^{\circ}$ of the sky. The passband is 600-1000 nm. The data are released in the Mikulski Archive for Space Telescopes (MAST) database. Similar to exoplanet transits, TESS is also very good at searching for binary eclipses. For KM UMa, two-minute cadence light curves are available in MAST for sectors 22 and 49. The fluxes are converted to magnitudes using the formula $\Delta mag=-2.5\times \log(flux)$ and then their mean values are subtracted. The phase-folded light curves are shown in Figure \ref{fig1}. The minimum shows a flat bottom, which means it should be a totally eclipsing binary. The O'Connell effect can be seen at both maxima and evolves with time, which may be due to spot activity.

SuperWASP is equipped with two telescopes located at the Roque de los Muchachos Observatory and the South African Astronomical Observatory. Each instrument has a field of view of approximately 482 square degrees, with an angular scale of 13.$^{''}$7 per pixel. Their primary science goals are to detect exoplanetary transits, monitor asteroids and comets, and discover optical transients. From the photometric data of SuperWASP, a significant flare activity was observed on 2007 February 28, as shown in the lower-right panel of Figure \ref{fig1}.

\begin{table}[ht!]
	\centering
	\caption{Photometric data of YNOs} 
		\begin{tabular}{lcccc}
			\hline \hline
			HJD	&	$\Delta$mag	&	Band	&	Ref	\\
            \hline
			2455261.20022 	&	-0.485 	&	R	&	60cm	\\
            2455261.20122 	&	-0.480 	&	R	&	60cm	\\
            2455261.20223 	&	-0.480 	&	R	&	60cm	\\
            $\cdots$ &$\cdots$	&$\cdots$ &	$\cdots$ \\
            2460419.33290 	&	-0.532 	&	I	&	70cm	\\
            2460419.33447 	&	-0.532 	&	I	&	70cm	\\
            2460419.33604 	&	-0.530 	&	I	&	70cm	\\
			\hline
		\end{tabular}
       \\ (This table is available in its entirety in machine-readable form.)
	\label{table_lc}
\end{table}

\subsection{Spectroscopy} 
We carried out medium-resolution spectroscopic observations of KM UMa on 2023 January 7 and 2023 January 29 using the 2.4 m telescope at the Thai National Observatory (TNO), National Astronomical Research Institute of Thailand (NARIT). The telescope is equipped with an ARC 4K CCD camera and a medium-resolution spectrograph. The wavelength range of the spectrometer is 3800-9000 {\AA} with a resolution of $R \sim 18,000$. The raw spectra were processed using the IRAF software, including spectral extraction, calibration, and normalization. Finally, six normalized spectra were obtained.

LAMOST, also known as the Guo Shoujing Telescope, located at the Xinglong Station in Hebei, China, has a large aperture and a wide field of view. The focal plane is equipped with 4000 fibers, allowing it to acquire 4000 spectra simultaneously. LAMOST has released more than 20 million low- and medium-resolution spectra to date. The wavelength range of the low-resolution spectrum is 3700-9000 {\AA}  with a resolution of $R \sim 1800$. The medium-resolution spectrum consists of the blue band (4950-5350 \AA) and the red band (6300-6800 \AA) with a resolution of $R \sim 7500$.  From the LAMOST Data Release (DR) 9 database, one low-resolution spectrum and hundreds of medium-resolution spectra were obtained, of which the high-signal-to-noise ratio (SNR) ones will be used for subsequent analyses.

 The Apache Point Observatory Galactic Evolution Experiment \citep[APOGEE;][]{2017AJ....154...94M} is a large-scale, stellar spectroscopic survey conducted in the near-infrared band ($1.51-1.70\  \mu m$). It is  one of the programs in the SDSS III. At the end of SDSS-III, APOGEE-1 collected 150,000 high-resolution ($R \sim 22,500$), and high-SNR ($\textgreater 100$) near-infrared spectra of 146,000 stars. Subsequently, the second generation of APOGEE was carried out in SDSS-IV and the survey was extended to the southern hemisphere \citep{2017AJ....154...28B}. In the DR 17, APOGEE-2 has observed 657,000 unique targets. APOGEE spectra are of three types: individual visit spectra (apVisit/asVisit), combined spectra (apStar/asStar), and pseudocontinuum normalized spectra (aspcapStar). For KM UMa, four apVisit-type spectra were found in the DR 17 database.

\section{Orbital Period Analysis} \label{section3}
\begin{figure}[ht!]
\plotone{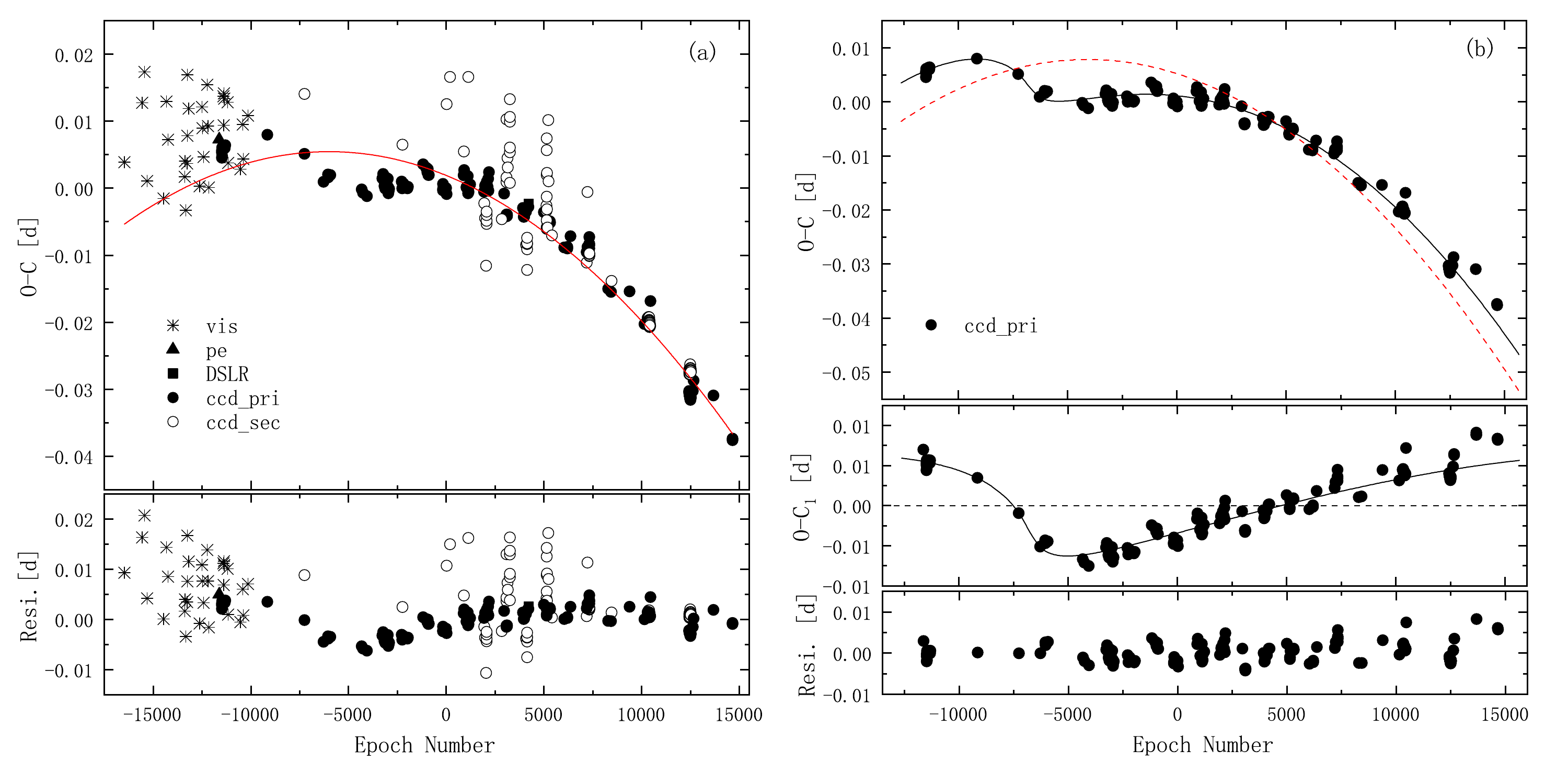}
\caption{O-C diagram of the eclipse times.
\label{fig2}}
\end{figure}

In order to study the variation of the orbital period, some eclipse times were collected from the O-C gateway. We searched the original literature of the data sources and removed data with unclear sources, uncertain accuracy, and obvious errors. Photometric data from the TESS, SuperWASP, and  the American Association of Variable Star Observers (AAVSO\footnote{\url{https://www.aavso.org/databases}}) database were also used to calculate some of the light minimum times. Combined with the data obtained from our observations, all times of minimum are listed in Table \ref{table_ETV}. The O-C values were calculated with the following linear ephemeris formula:
\begin{equation}
 Min.I(HJD)=2455261.25102+0_{.}^{d}3518608 \times E,
\label{euqation1}
\end{equation}
where $HJD_{0}$ is one of the eclipse times obtained in the present paper and the period is from the International Variable Star Index (VSX) catalog. The O-C values versus the epoch number \textit{E} are displayed in panel (a) of Figure \ref{fig2}. The overall trend of the O-C curve presents a downward parabolic variation. The weights of visual data, CCD secondary minima, and high-precision primary minima (CCD, Pe, and  DSLR) are assigned as 1, 5, and 10, respectively. Then, the following equation is derived based on the weighted least-squares method:
\begin{equation}\label{equation2}
\begin{aligned}
 Min.I(HJD)=2455261.25292(\pm0.00024)+0_{.}^{d}35185961578\\
 (\pm0.000000028)
 \times E
 -9.87131(\pm0.30683)\times10^{-11}\times E^2.
\end{aligned}
\end{equation}
As can be seen in Figure \ref{fig2} (a), there is a large dispersion in the secondary minima times. This may be due to the shallower amplitude, which is more influenced by the spot activity \citep{2013ApJ...774...81T,2015MNRAS.448..429B,2021RMxAA..57..335Y}. From the residuals, there are also some trends in the primary minima times. To obtain more precise results, we focus on the primary minima O-C variations, as shown in panel (b) of Figure \ref{fig2}. The trend in Figure \ref{fig2} (b) may be due to the light travel time effect (LTTE) of a third body. Then, the following general equation was used to fit the primary minima O-C curve \citep{1952ApJ...116..211I}:
\begin{equation}
\begin{aligned}
 O-C=&\Delta T_0 +\Delta P_0 E+\frac{\beta}{2} E^2\\
 &+A[(1-e_{3}^{2})\frac{\sin(\nu+\omega)}{1+e_{3}\cos{\nu}}+e\sin{\omega}]\\
 =&\Delta T_0 +\Delta P_0 E+\frac{\beta}{2} E^2\\ 
 & +A[\sqrt{1-e_3^2} \sin E^* \cos \omega+\cos E^* \sin \omega].
\end{aligned}
\end{equation}
From Kepler's equation:
\begin{equation}
M=E^*-e_3 \sin E^*=\frac{2 \pi}{P_3}\left(t-T_3\right)
\end{equation}
Here, $A=a_{12}\sin{i_3}/c$ is the semiamplitude of the LTTE, $a_{12}$  is the semi-major axis of the eclipsing pair, $e_{3}$ is the eccentricity, $\nu$ is the true anomaly, $\omega$ is the longitude of periastron, and $E^{*}$ is the eccentric anomaly. $M$, $T_3$, and $P_3$ are the mean anomaly, time of periastron passage, and period of the third body, respectively. Then, LTTE fitting was performed with the help of the Python $emcee$ package using the Markov Chain Monte Carlo (MCMC) method. The results of the parameter fitting are displayed in Table \ref{table_3body_para}, and the MCMC corner plots are shown in Figure \ref{corner} in the Appendix. It implies the possible presence of a third body orbiting the common center of mass with the inner pair. Given that the CCD primary epoch spans only 25 yr, additional data are required to improve the accuracy of parameter fitting.

\begin{table}[htbp] 
	\centering
	\caption{The light minimum times of KM UMa}
	
	\begin{tabular}{ccccc|ccccc}
		\hline
		Eclipse Timings	&	Error	&	\textit{E}	&	O-C		&	Reference	&	Eclipse Timings	&	Error	&	\textit{E}	&	O-C		&	Reference	\\
		(HJD)	&	(d)	&		&	(d)	&		&	(HJD)	&	(d)	&		&	(d)		&		\\
		\hline
    2449465.40380 	&		&	-16472	&	0.00388 	&	(1)	&	2458905.10152 	&	0.00016 	&	10356	&	-0.01995 	&	(42)	\\
    2449787.36530 	&		&	-15557	&	0.01275 	&	(1)	&	2458905.27756 	&	0.00020 	&	10356.5	&	-0.01984 	&	(42)	\\
    2449828.53760 	&		&	-15440	&	0.01733 	&	(1)	&	2458905.45336 	&	0.00017 	&	10357	&	-0.01997 	&	(42)	\\
    2449877.43000 	&		&	-15301	&	0.00108 	&	(1)	&	2458905.62929 	&	0.00019 	&	10357.5	&	-0.01997 	&	(42)	\\
    2450170.52740 	&		&	-14468	&	-0.00157 	&	(1)	&	2458905.80524 	&	0.00016 	&	10358	&	-0.01995 	&	(42)	\\
    2450225.43220 	&		&	-14312	&	0.01295 	&	(1)	&	2458905.98111 	&	0.00019 	&	10358.5	&	-0.02001 	&	(42)	\\
        $\cdots$	&	$\cdots$	&	$\cdots$	&	$\cdots$	&	$\cdots$	&	$\cdots$	&	$\cdots$	&	$\cdots$	&	$\cdots$	&	$\cdots$	\\
    2458904.04604 	&	0.00018 	&	10353	&	-0.01984 	&	(42)	&	2459664.23365 	&	0.00014 	&	12513.5	&	-0.02749 	&	(42)	\\
    2458904.22215 	&	0.00017 	&	10353.5	&	-0.01967 	&	(42)	&	2459702.05590 	&		&	12621	&	-0.03028 	&	(43)	\\
    2458904.39783 	&	0.00016 	&	10354	&	-0.01992 	&	(42)	&	2459715.42815 	&	0.00007 	&	12659	&	-0.02874 	&	(10)	\\
    2458904.57399 	&	0.00016 	&	10354.5	&	-0.01969 	&	(42)	&	2460076.43512 	&	0.00011 	&	13685	&	-0.03095 	&	(10)	\\
    2458904.74978 	&	0.00017 	&	10355	&	-0.01983 	&	(42)	&	2460416.32621 	&	0.00015 	&	14651	&	-0.03740 	&	(44)	\\
    2458904.92581 	&	0.00016 	&	10355.5	&	-0.01973 	&	(42)	&	2460419.14085 	&	0.00015 	&	14659	&	-0.03763 	&	(44)	\\
		\hline
	\end{tabular}
\begin{flushleft}
            \footnotesize
            \textbf{Note.}References: (1)\citet{2001GEOCE..26.....V}; (2)\citet{1999IBVS.4810....1E}; (3)o-c gateway (4)BBSAG NO.125; (5)VSOLJ NO.42; (6)\citet{2005IBVS.5603....1D}; (7)SuperWASP; (8)YNO 1 m; (9)\citet{2009IBVS.5874....1H};
             (10)AAVSO;  (11)VSOLJ NO.48; (12)\citet{2010IBVS.5918....1H}; (13)\citet{2010IBVS.5938....1D}; (14)\citet{2011IBVS.5959....1H}; (15)\citet{2011IBVS.5974....1D}; (16)YNO 60 cm; (17) \citet{2011IBVS.5966....1N}; (18)\citet{2010IBVS.5945....1D}; (19)VSOLJ NO.51; (20)VSOLJ NO.53; (21)\citet{2011IBVS.5992....1D}; (22)\citet{2013OEJV..160....1H}; (23)\citet{2012IBVS.6018....1N}; (24)\citet{2013IBVS.6070....1H}; (25)\citet{2013IBVS.6048....1H}; (26)VSOLJ NO.55; (27)\citet{2012IBVS.6029....1D}; (28)\citet{2013IBVS.6050....1N}; (29)VSOLJ NO.56; (30)\citet{2014OEJV..165....1H}; (31)\citet{2015OEJV..168....1H}; (32)VSOLJ NO.59; (33)\citet{2017IBVS.6196....1H}; (34)\citet{2015IBVS.6131....1N}; (35)\citet{2017OEJV..179....1J}; (36)\citet{2019JAVSO..47..106S}; (37)\citet{2019JAVSO..47..265S}; (38)VSOLJ NO.60; (39)\citet{2021OEJV..211....1L}; (40)VSOLJ NO.66; (41)VSOLJ NO.69; (42)TESS; (43)VSOLJ NO.108; (44)70cm\\
            (This table is available in its entirety in machine-readable form.)
            
        \end{flushleft}
    \label{table_ETV}
\end{table}

\begin{table}[htbp]
\centering
\caption{Orbital parameters of the third body.} 
\begin{tabular}{lccc}
\hline\hline
Parameter   & Value  & Unit     \\
\hline
Revised epoch, $HJD_0$&   $2455261.25623(4)$ & days\\
Revised period, $P_0$ &   $0.3518595138(19)$    & day \\
Long-term change of the orbital period, $\beta$ & $-3.149(8)\times 10^{-10}$ &d cycle$^{-1}$\\
Amplitude, $A$ & $0.01124(4)$  &  day\\
Projected semi-major axis,$a_{12}\sin{i_3}$  & $1.946(7)$  &  au\\
Eccentricity, $e_3$ & $ 0.805(2)$  &  -\\
Longitude of periastron, $\omega$ & $-157.054(207)$  & deg\\
Orbital period, $P_3$ & $33.66(12)$  &yr\\
Time of periastron passage, $T_3$ & $2452915(12).47306$ & days \\
Mass function, $f(m)$ & $0.0065(1)$  &$ M_{\odot}$\\
\hline 
\end{tabular}
\label{table_3body_para}
\end{table}

\section{Spectroscopic and Photometric Study}\label{section4}
\subsection{Radial velocities and equivalent widths}

\begin{figure}[ht!]
\plotone{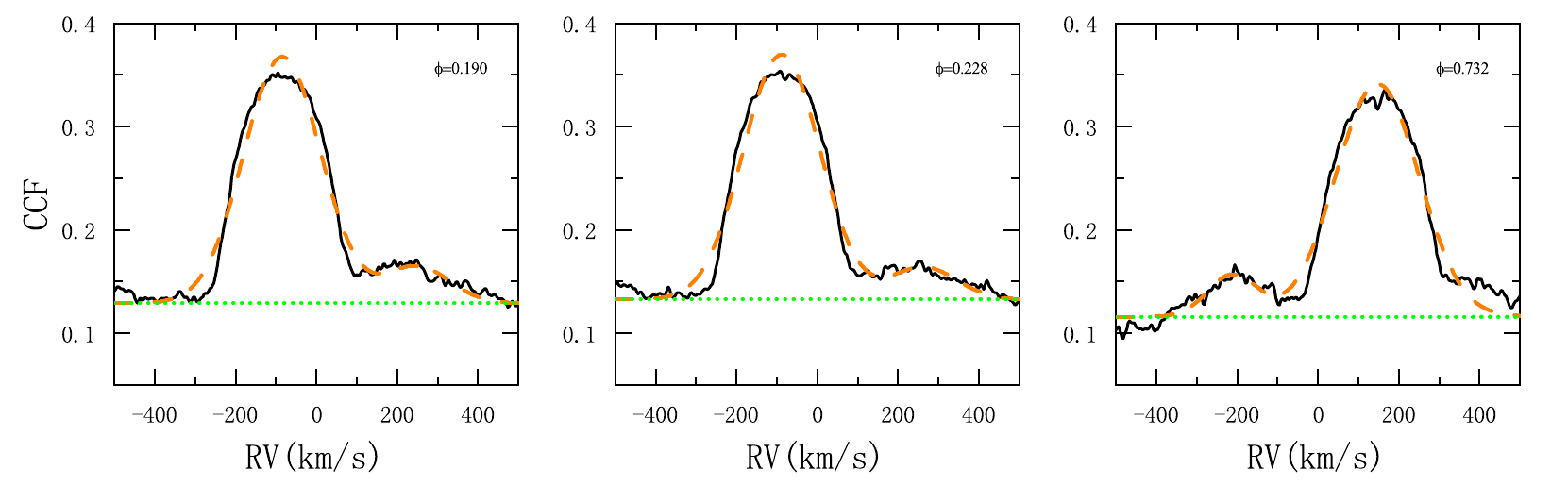}
\caption{CCF profiles of SDSS APOGEE spectra.
\label{CCF}}
\end{figure}

\begin{figure}[ht!]
\plotone{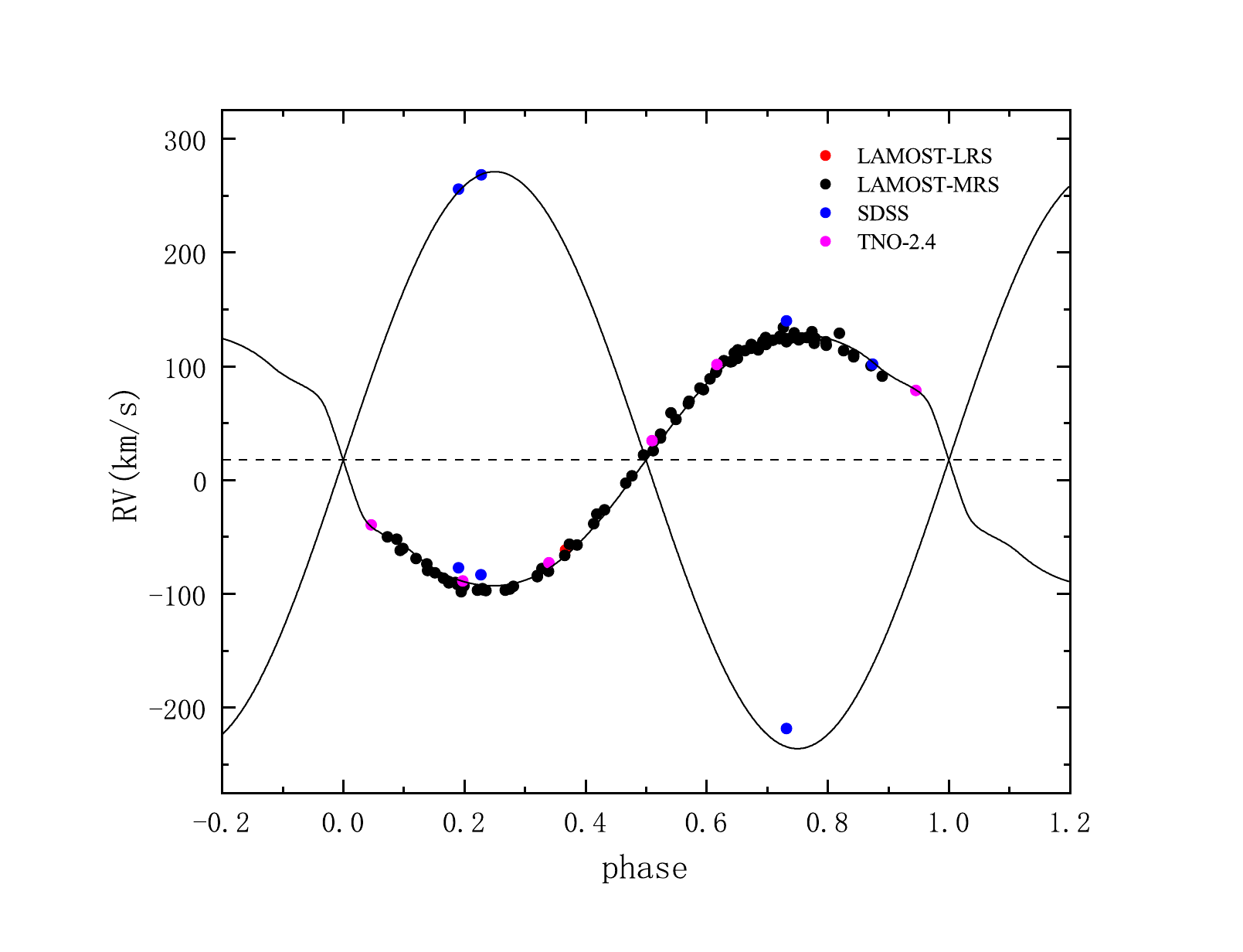}
\caption{The RV curves of KM UMa.
\label{RV}}
\end{figure}

To accurately determine the fundamental physical parameters of the system, the radial velocity (RV) was calculated using the cross-correlation function (CCF) method. The template spectrum adopted synthetic spectra based on the PHOENIX model of \citet{2013A&A...553A...6H}. All observed spectra were converted to vacuum wavelengths. For the spectra observed by LAMOST and TNO, only one component is visible in their CCF profiles due to the large luminosity difference between the two components in the optical band. Fortunately, the infrared spectra from APOGEE are able to exhibit the CCF profile of the secondary star, as illustrated in Figure \ref{CCF}. The profiles were fitted with single or double Gaussian functions, and the fitting results are shown in Table \ref{table_RV}. The uncertainties reported here are fitting errors and are expected to be larger in reality. Then, the RV data were fitted using the Wilson–Devinney (W-D) code \citep{1971ApJ...166..605W,2012AJ....144...73W,2014ApJ...780..151W}. The RV standard deviations for the primary and secondary components were set to 1 and 10, respectively. 
Finally, we obtain a mass ratio of $q = 0.45(\pm 0.04)$ and a center-of-mass velocity of $V_{\gamma} = 17.61\ (\pm 0.25)$  km s$^{-1}$. The actual error may be a bit larger due to less RV phase coverage of the secondary component. The phase-folded RV curves and the best-fit theoretical curves are shown in Figure \ref{RV}.

\begin{figure}[ht!]
\plotone{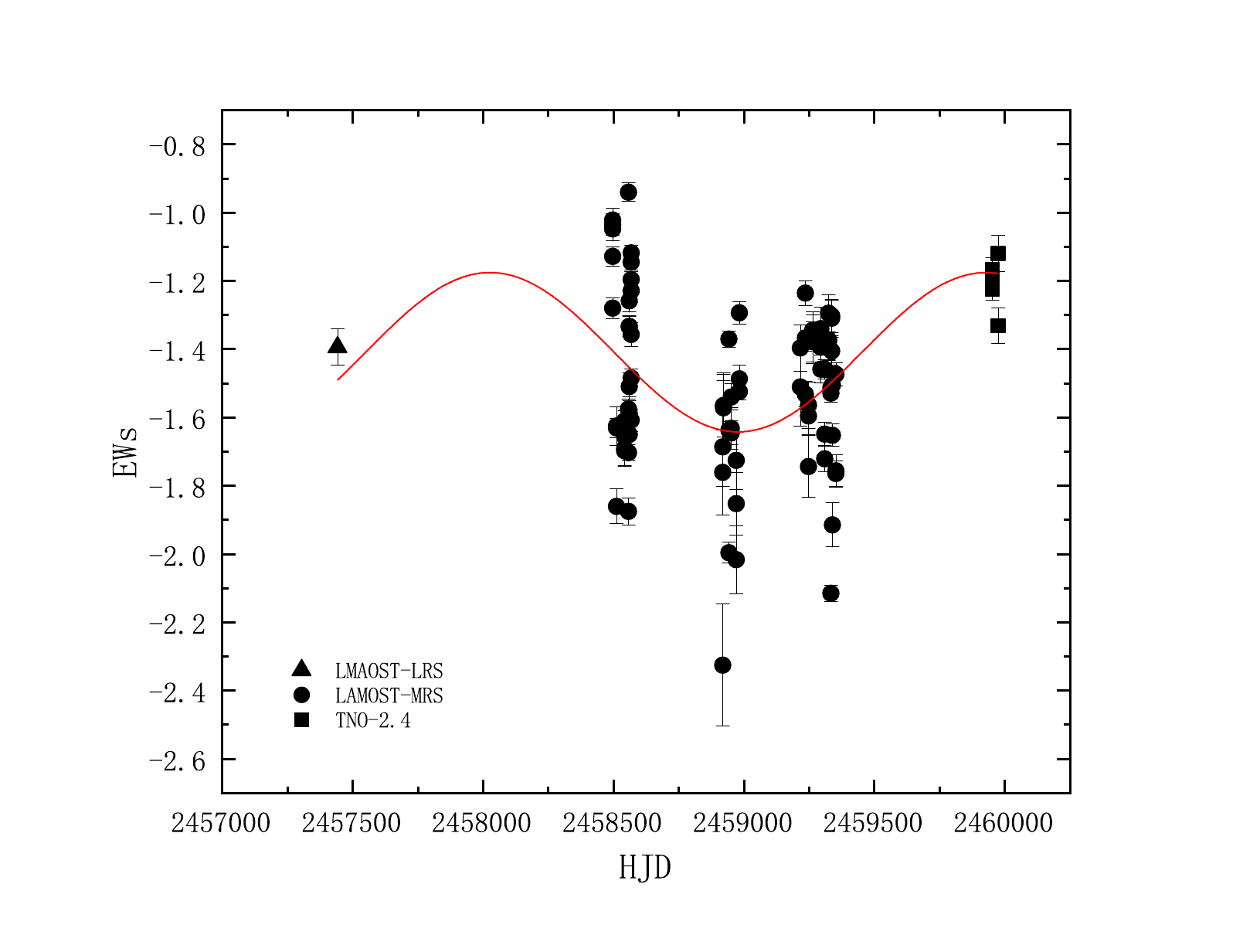}
\caption{The EW variation of the $H_{\alpha}$ line. The red line represents a sinusoidal fit with a period of $P=5.21\ (\pm 0.67)$ yr.
\label{EW}}
\end{figure}

In the optical band, the spectra are dominated by the primary component, hence the spectral lines essentially correspond to variations in the primary. The magnetic activity level of the primary component was measured by calculating the equivalent widths (EWs) of the $H_\alpha$ line in the LAMOST and TNO spectra. The variation of EW over time is depicted in Figure \ref{EW}. From the figure, a periodic variation is observed, and a sinusoidal fit reveals a period of approximately $P=5.21\ (\pm 0.67)$ yr, which may be attributed to magnetic activity cycles.

\subsection{Light-curve Modeling}
\begin{figure}[htbp]
    \centering
    \begin{minipage}{1\textwidth}
        \centering
        \includegraphics[scale=0.4]{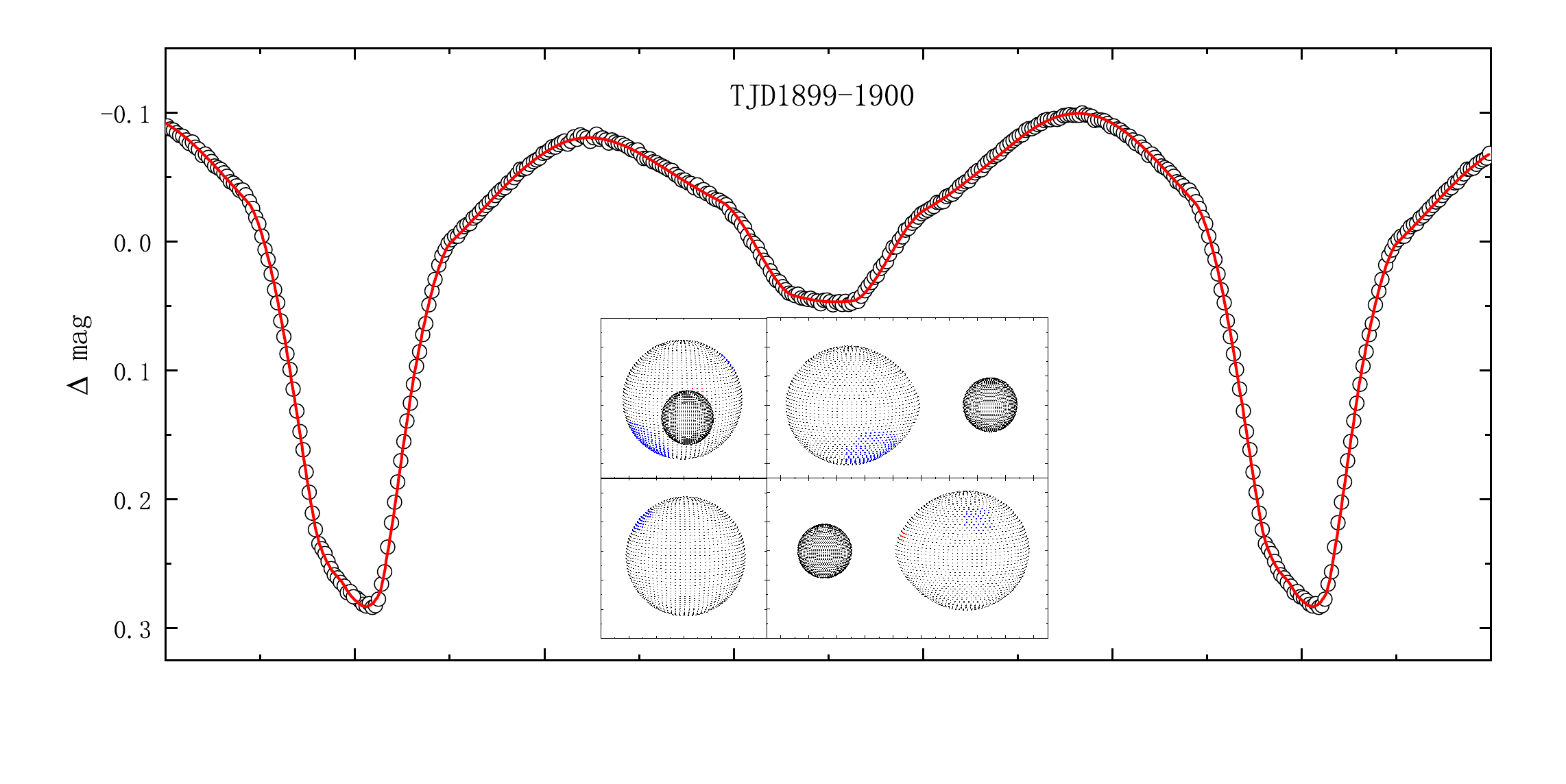} 
    \end{minipage}
    
    \vspace{-1.5cm} 
    
    \begin{minipage}{1\textwidth}
        \centering
        \includegraphics[scale=0.4]{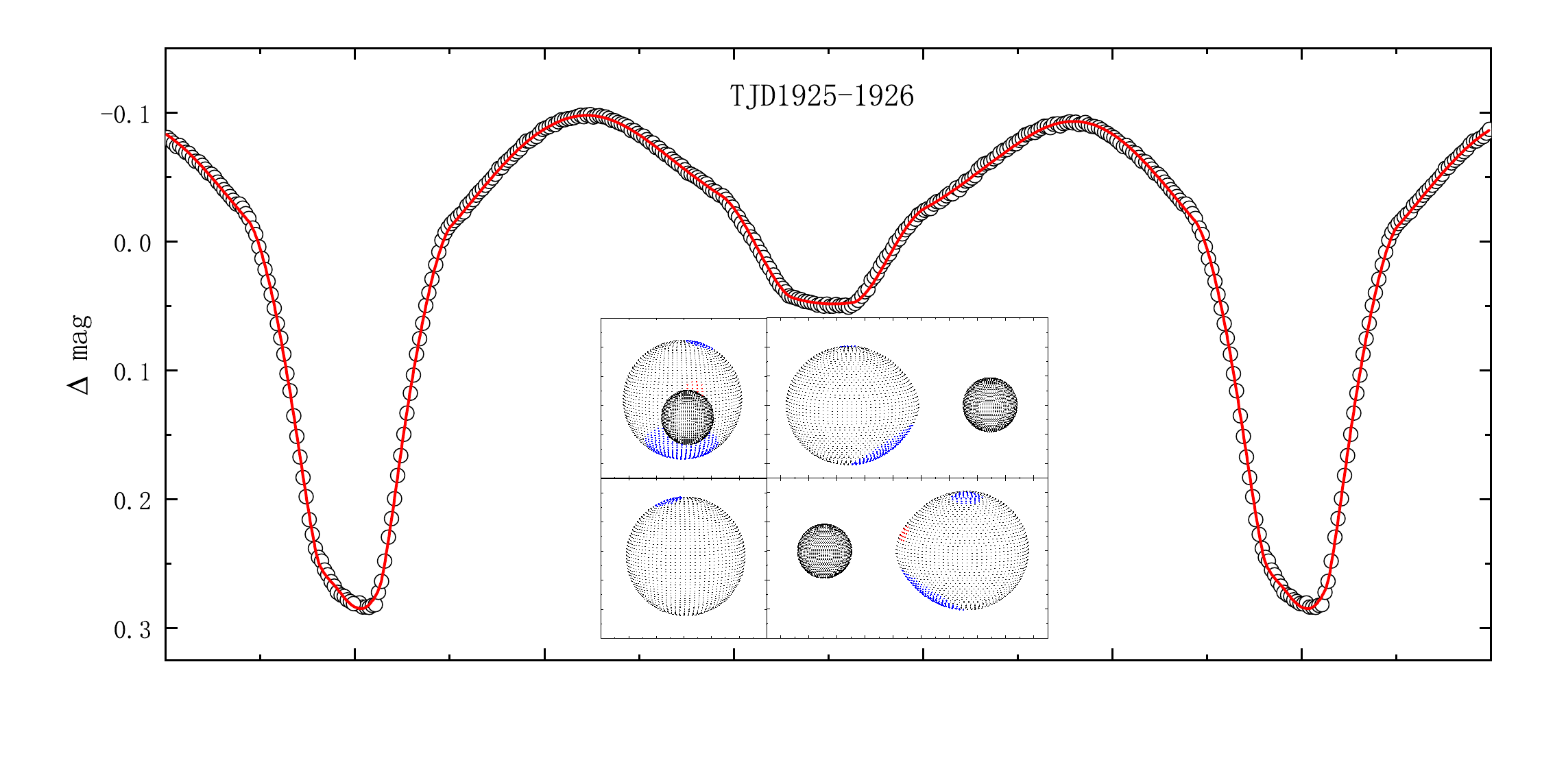} 
    \end{minipage}
    
    \vspace{-1.5cm} 
    
    \begin{minipage}{1\textwidth}
        \centering
        \includegraphics[scale=0.4]{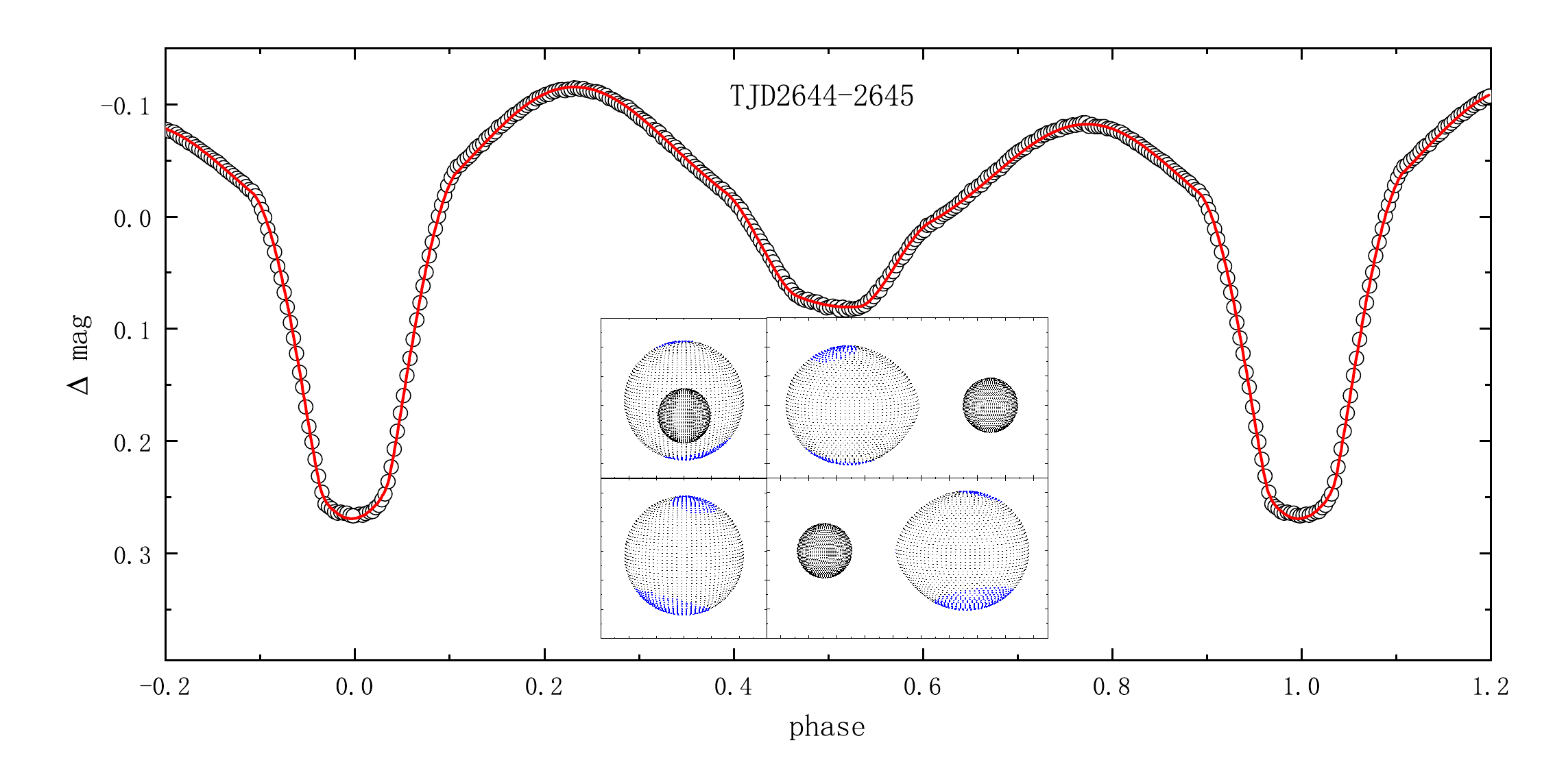} 
    \end{minipage}
    \caption{Light curves (black open circles) at three different stages, along with their theoretical fitting curves (red solid line) and geometric configurations. The blue dots represent dark spots, and the red dots represent hot spot or stellar facula.}
    \label{WD}
\end{figure}

The TESS photometric data from two sectors were analyzed using the W-D program. Based on the variation in the maxima, three segments of data were selected for analysis, each covering approximately three cycles. To minimize the impact of starspot activity, the initial analysis used relatively symmetric light curves (TJD 1925-1926) to determine the mode and basic solution. Based on the stellar atmospheric parameters from LAMOST, the surface temperature of the primary component was set to 5353 K. The gravity-darkening coefficient $g_1 = g_2 = 0.32$ and the bolometric albedo $A_1 = A_2 = 0.5$  were adopted for the convective envelope \citep{1967ZA.....65...89L,1969AcA....19..245R}. After testing modes 2, 4, and 5, it was determined that KM UMa is a detached system, i.e., mode 2 is the most likely solution. Then, we fixed the mass ratio to $q = 0.45$ and selected mode 2 for further analysis of the three segments of data. For sector 22, there are significant distortions at both primary and secondary minima. Additionally, Max I (phase=0.25) is fainter than Max II (phase=0.75), but at the secondary minimum, the left edge during eclipse is brighter than the right edge. This indicates the presence of at least three spots in the system, affecting the light variations during the primary minimum, secondary minimum, and outside of eclipses. Furthermore, from the multicolor light curves in Figure \ref{fig1}, the primary minimum appears relatively symmetric in the \textit{I} band but distorted in the \textit{B} band, suggesting that the spot radiates primarily at the blue band. Moreover, there is an increase in brightness at the left edge of the primary minimum, indicating the presence of a hot spot on the stellar surface, most likely on the primary one. Since the multicolor light curves at this time closely resemble the light variation in sector 22, we assume that there are two dark spots and one hot spot on the surface of the primary in sector 22. The hot spot on the primary star's surface may correspond to solar facula. After adjusting the parameters several times, the final fitting results were obtained. The parameters of the fitting results are listed in Table \ref{table_WD}, and the model is depicted with a red curve in Figure \ref{WD}.

\begin{table}[htbp]
\centering
\caption{Photometric solutions for KM UMa.} 
\begin{tabular}{lcccc}
\hline\hline
Parameters & TJD 1899-1900    &TJD 1925-1926      & TJD 2644-2645 \\
\hline
Mode  &  \multicolumn{3}{c}{Detached binaries (mode 2)}	\\
$q (M_2/M_1)$ & 0.45(fixed)  & 0.45(fixed)  &0.45(fixed)\\
$i (^{\circ})$ & 82.951(71) &  83.05(12)  &   83.984(81)        \\
$T_1$ (K)  &5353(fixed)   &   5353(fixed) &5353(fixed)\\
$T_2$ (K)  &3643(17)     &  3673(5)  & 3635(10) \\
$\Omega_1$ &2.8626(25)  &  2.8708(22)      &2.8672(24) \\
$\Omega_2$ & 3.6848(74)    & 3.6733(75)       &3.6638(56) \\
$L_1 /(L_1+L_2)_{\rm{TESS}}$ & 0.968541(16)  & 0.967982(41)  &0.970377(19) \\
$L_2 /(L_1+L_2)_{\rm{TESS}}$ &  0.031459(16)     & 0.032018(41)  &  0.029623(19)\\
$r_1$(pole)  & 0.40902(41)  & 0.40749(36)       & 0.40803(39) \\
$r_1$(point)  & 0.4974(13)   & 0.4928(10)       &0.4944(12)\\
$r_1$(side)  & 0.43237(52)   &   0.43046(45)      & 0.43114(49)\\
$r_1$(back)  & 0.45513(64)     &0.45277(55)        &0.45360(60) \\
$r_2$(pole)  & 0.18573(56)    & 0.18268(35)       &0.18705(43)\\
$r_2$(point)  &0.19289(66)    & 0.19371(67)       &0.19444(51)\\
$r_2$(side)  & 0.18768(58)    &  0.18842(60)      &0.18906(45) \\
$r_2$(back)  & 0.19160(63)    &  0.19240(65)      & 0.19310(49) \\
$f_1 (\%)$  & 0.8922(20)    & 0.8819(17)       &0.8862(18)\\
$f_2 (\%)$  & 0.2185(11)    &  0.2218(11)      &0.22442(85)\\
\hline 
Spot1  &       &        &\\
Colatitude (radian) & 2.33301      & 2.50119       &2.81946\\
Longitude (radian) & 5.22440      &  6.21728      &2.15098\\
Radius (radian) & 0.43839      & 0.63245      &0.69334\\
Temp.factor&  0.93053     &  0.95416      &0.79804\\
\hline 
Spot2  &       &        &\\
Colatitude (radian) &  0.90426  & 0.31030        &0.33749\\
Longitude (radian) & 1.80368 &  1.63341      &3.69546\\
Radius (radian) & 0.21737  &  0.22883      &0.37230 \\
Temp.factor& 0.91365 &  0.67921      &0.80430\\
\hline 
Spot3  &       &        &\\
Colatitude (radian) &  1.34375  & 1.30176        &\\
Longitude (radian) &0.21026&  0.16477     &\\
Radius (radian) & 0.07645 & 0.09909     &\\
Temp.factor&1.58813  &  1.48652       &\\
\hline 
\end{tabular}
\label{table_WD}
\end{table}

\section{Discussions and Conclusions}\label{section5}
Based on photometric and spectroscopic analysis, KM UMa is a short-period detached eclipsing binary with a period of approximately 0.35186 day and a mass ratio of $q=0.45\ (\pm0.04)$. The luminosity of the two components differs significantly, with the primary star contributing approximately 97$\%$. The volume filling factor of the primary component is approximately $89\%$, and that of the secondary component is approximately $22\%$. The absolute parameters of the binary system were calculated by combining the results of spectroscopic orbit and light-curve modeling, which are listed in Table \ref{table_ab_para}.

Compared to other detached eclipsing binaries, KM UMa has a relatively short period. According to dynamo theory, it is expected to exhibit higher activity. In fact, its light curves show a significant O'Connell effect.  From the SuperWASP photometric data, we identified a superflare event occurring on 2007 February 28. Using the absolute parameters from Table \ref{table_ab_para} and applying the method of \citet{2013ApJS..209....5S}, we estimated the flare energy to be roughly $5\times 10^{34}$ erg assuming it occurred on the primary component. The variation in the EWs of the $H_{\alpha}$ line in LAMOST and TNO spectra indicates the presence of a $5.21\ (\pm0.67)$ yr magnetic activity cycle. The multiband light curves and W-D analysis indicate the presence of spots on the surface of the primary component. All of the above phenomena indicate that KM UMa is an active system. 

The orbital period variations were analyzed using the O-C method, as shown in Figure \ref{fig2} (a). The general (O-C) trend shows a downward parabolic variation, indicating that the orbital period is decreasing at a rate of  $d P / d t=-2.05 \times 10^{-7}$ days yr ${ }^{-1}$, possibly due to angular momentum loss caused by magnetic braking. In order to study the orbital period variations in more detail, the high-precision CCD primary eclipse timing variations are shown in panel (b) of Figure \ref{fig2}. 
Based on the continuous variation in the shape of the O-C data, we are more inclined to attribute this variation to the LTTE of a third body rather than to a magnetic activity cycle, which is the same as the statistical view of \citet {2010MNRAS.405.1930L}. The presence of a third body may accelerate the evolution of the inner binary, leading to such a short orbital period by extracting angular momentum from the system\citep{2006AJ....131.3028Q,2007AJ....133..357Q}. By using the equation, 
\begin{equation}
f(m)=\frac{\left(M_3 \sin i\right)^3}{\left(M_1+M_2+M_3\right)^2}=\frac{4 \pi^2}{G T^2} \times\left(a_{12} \sin i\right)^3,
\end{equation}
the mass function is calculated as $f(m)=0.0065$ $M_{\odot}$. The minimum mass of the third object is derived as $M_3=0.31$ $M_{\odot}$ at $i_3=90^{\circ}$. Assuming it is a main-sequence star, its luminosity is about 0.01 $L_{\odot}$ \citep{2013ApJS..208....9P}, which accounts for 1$\%$ of the total luminosity of the system. Since the light curve is strongly influenced by magnetic activity and TESS has only a single band, the third light solution was not considered. More high-precision photometric and spectroscopic data are needed to validate its existence.

In addition, a visual companion (Vis) near KM UMa was found through the Gaia DR2 and DR3 databases \citep{2018A&A...616A...1G,2023A&A...674A...1G}. Table \ref{table_parallaxes} provides the parallaxes of KM UMa and Vis in the Gaia DR3 database, which are essentially the same. Moreover, their proper motions are also very close to each other in both the R.A. and Decl. directions. All astrometric parameters indicate that Vis and KM UMa form a gravitationally bound system,  similar to XZ CMi \citep{2020RAA....20..133W}.  Based on the angular distance and parallax, it is derived that Vis is about 2,178 au from the binary pair. Combined with the O-C analysis, KM UMa is probably in a 2+1+1 hierarchical quadruple system. The short-period active detached binary in a quadruple system makes KM UMa a good astrophysical laboratory for the study of the formation and evolution of multiple systems.

\begin{table}[ht!]
	\centering
	\caption{Absolute parameters of KM UMa} 
		\begin{tabular}{lcc}
			\hline \hline Parameters & Star 1 & Star 2 \\
			\hline Mass $\rm M_{\odot}$ & $1.27(\pm0.14)$ & $0.57(\pm0.06)$ \\
			Radius $\rm R_{\odot}$ & 1.15($\pm0.04$) & 0.49($\pm0.02$) \\
			Luminosity $\rm L_{\odot}$ & 0.97($\pm0.06$) & 0.038($\pm0.003$) \\
			Semi-major axis $\rm R_{\odot}$ & \multicolumn{2}{c}{2.57($\pm0.07$)} \\
			\hline
		\end{tabular}
	\label{table_ab_para}
\end{table}

\begin{table}[htbp]
\centering
\caption{Astrometric Parameters of KM UMa and the visual companion.} 
\begin{tabular}{lccccc}
\hline\hline
Name & Coordinate & Parallax (mas)  &pm.ra (mas yr$^{-1}$) & pm.dec (mas yr$^{-1}$)\\
\hline
KM UMa& 11:47:49.04 +35:13:35.23   &9.16(2) &-78.68(1) &-6.48(1)\\
Visual companion & 	11:47:50.65 +35:13:38.44 &9.15(3) &-77.93(2) & -5.51(2)\\
\hline 
\end{tabular}
\label{table_parallaxes}
\end{table}

\section*{Acknowledgments}
This work was supported by the International Cooperation Projects of the National Key R$\&$D Program (No. 2022YFE0127300), the Science Foundation of Yunnan Province (No. 202401AS070046), the International Partership Program of Chinese Academy of Sciences (No. 020GJHZ2023030GC), the Agency of innovative development under the Ministry of Higher Education,Science and Innovation of the Republic of Uzbekistan, Project (AL-5921122128), CAS'Light of West China' Program, and Yunnan Revitalization Talent Support Program. New CCD photometric data were obtained with the Sino-Thai 70 cm telescope in Lijiang and 60 cm and 1m telescopes at Yunnan Observatories. The TESS photometric data were downloaded from the MAST database. The spectral data of KM UMa were provided by LAMOST, SDSS, and the TNO 2.4 m telescope. We acknowledge the support of the staff of the TNO 2.4 m telescope. This work also makes use of data from SuperWASP, AAVSO, and Gaia. The authors thank these teams for providing public data.

\appendix
\section*{Appendix information}
In this section, we present the RV measurements in Table \ref{table_RV} and the corner plot of the LTTE fitting results in Figure \ref{corner}.

\renewcommand\thefigure{A1}
\renewcommand\thetable{A1}

\begin{figure}[ht!]
\plotone{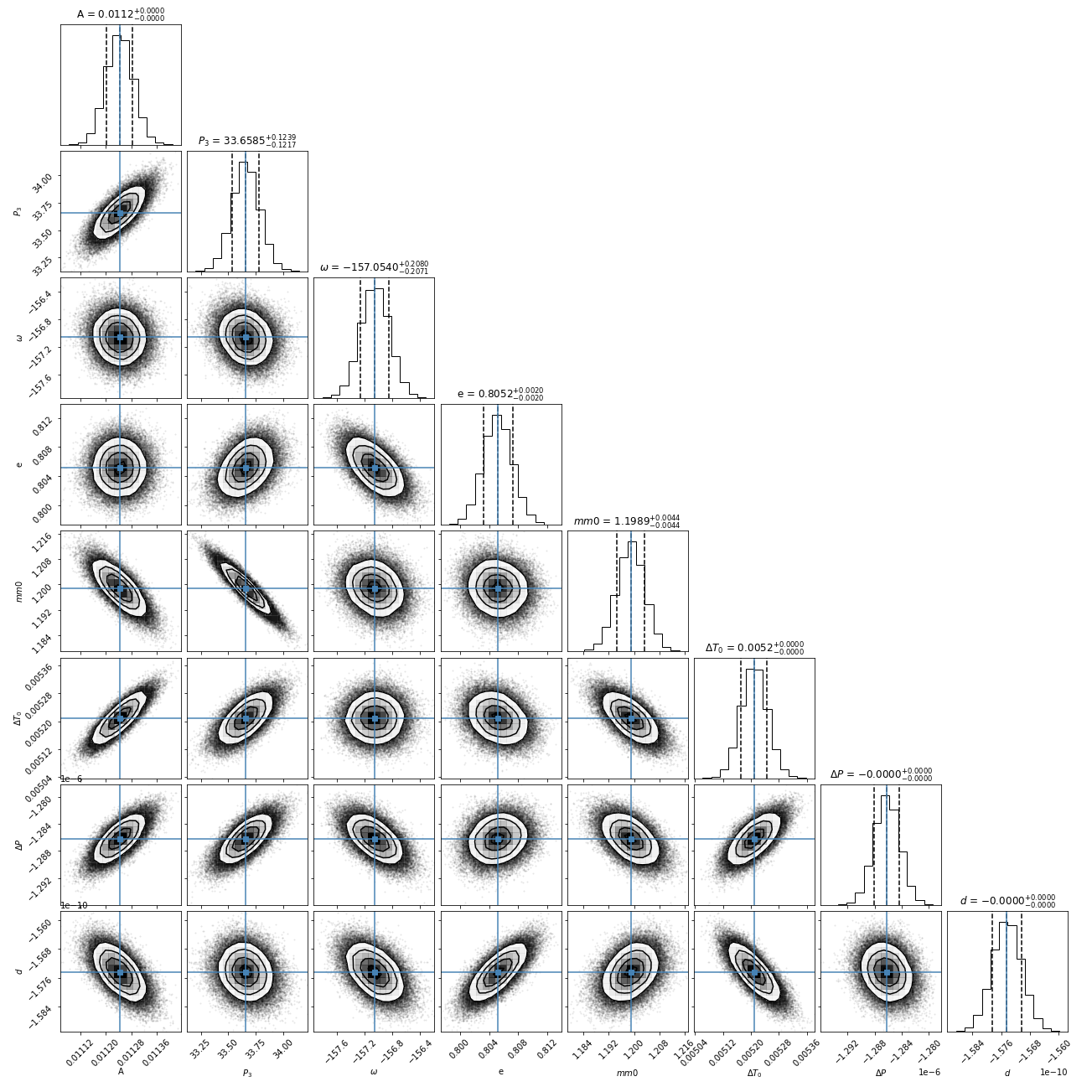}
\caption{Corner plots of MCMC fitted O-C curves. Here, $d=\beta/2$ and $mm0=2\pi(HJD_0-T_3)/P_3$.
\label{corner}}
\end{figure}

\begin{table}[htbp] 
	\centering
	\caption{KM UMa RV measurements.}
	
	\begin{tabular}{ccccc|ccccc}
		\hline
HJD	&	Phase	&	RV	&	Error	&	Resource	&	HJD	&	Phase	&	RV	&	Error	&	Resource	\\
	&		&	(km/s)	&	(km/s)	&		&		&		&	(km/s)	&	(km/s)	&		\\

		\hline
2457443.26091 	&	0.367 	&	-61.30 	&	0.46 	&	LAMOST-LRS	&	2459216.42325 	&	0.423 	&	121.54 	&	0.89 	&	LAMOST-MRS	\\
2458496.35802 	&	0.358 	&	-77.81 	&	0.77 	&	LAMOST-MRS	&	2459216.43922 	&	0.439 	&	110.54 	&	0.90 	&	LAMOST-MRS	\\
2458496.37399 	&	0.374 	&	-56.37 	&	0.81 	&	LAMOST-MRS	&	2459236.27943 	&	0.279 	&	-95.53 	&	0.79 	&	LAMOST-MRS	\\
2458496.38997 	&	0.390 	&	-29.80 	&	0.85 	&	LAMOST-MRS	&	2459236.29540 	&	0.295 	&	-95.74 	&	0.80 	&	LAMOST-MRS	\\
2458496.40663 	&	0.407 	&	-2.58 	&	0.82 	&	LAMOST-MRS	&	2459236.31138 	&	0.311 	&	-84.79 	&	0.77 	&	LAMOST-MRS	\\
2458496.42261 	&	0.423 	&	26.01 	&	0.74 	&	LAMOST-MRS	&	2459246.30632 	&	0.306 	&	134.17 	&	1.45 	&	LAMOST-MRS	\\
2458512.31727 	&	0.317 	&	114.61 	&	0.84 	&	LAMOST-MRS	&	2459246.32299 	&	0.323 	&	130.59 	&	1.18 	&	LAMOST-MRS	\\
2458512.33325 	&	0.333 	&	124.34 	&	0.82 	&	LAMOST-MRS	&	2459246.33896 	&	0.339 	&	129.21 	&	1.22 	&	LAMOST-MRS	\\
2458512.34991 	&	0.350 	&	120.37 	&	0.82 	&	LAMOST-MRS	&	2459247.33553 	&	0.336 	&	114.59 	&	0.94 	&	LAMOST-MRS	\\
2458536.24234 	&	0.242 	&	116.56 	&	0.87 	&	LAMOST-MRS	&	2459247.35151 	&	0.352 	&	125.20 	&	0.99 	&	LAMOST-MRS	\\
2458542.24866 	&	0.249 	&	123.17 	&	0.82 	&	LAMOST-MRS	&	2459247.36817 	&	0.368 	&	129.56 	&	0.92 	&	LAMOST-MRS	\\
2458542.26463 	&	0.265 	&	118.65 	&	0.82 	&	LAMOST-MRS	&	2459264.26939 	&	0.269 	&	124.54 	&	0.85 	&	LAMOST-MRS	\\
2458542.28060 	&	0.281 	&	108.33 	&	0.85 	&	LAMOST-MRS	&	2459264.28605 	&	0.286 	&	113.94 	&	0.90 	&	LAMOST-MRS	\\
2458542.29727 	&	0.297 	&	91.25 	&	0.85 	&	LAMOST-MRS	&	2459264.30202 	&	0.302 	&	100.39 	&	0.96 	&	LAMOST-MRS	\\
2458558.20064 	&	0.201 	&	-51.78 	&	0.81 	&	LAMOST-MRS	&	2459295.18592 	&	0.186 	&	111.88 	&	0.92 	&	LAMOST-MRS	\\
2458558.21800 	&	0.218 	&	-73.75 	&	0.73 	&	LAMOST-MRS	&	2459295.20258 	&	0.203 	&	121.88 	&	0.89 	&	LAMOST-MRS	\\
2458558.23467 	&	0.235 	&	-90.01 	&	0.71 	&	LAMOST-MRS	&	2459295.21925 	&	0.219 	&	125.02 	&	0.84 	&	LAMOST-MRS	\\
2458558.25064 	&	0.251 	&	-96.36 	&	0.69 	&	LAMOST-MRS	&	2459309.09444 	&	0.094 	&	-90.22 	&	0.72 	&	LAMOST-MRS	\\
2458561.16864 	&	0.169 	&	40.50 	&	0.77 	&	LAMOST-MRS	&	2459309.11111 	&	0.111 	&	-96.89 	&	0.66 	&	LAMOST-MRS	\\
2458561.18531 	&	0.185 	&	69.37 	&	0.77 	&	LAMOST-MRS	&	2459309.12708 	&	0.127 	&	-96.81 	&	0.74 	&	LAMOST-MRS	\\
2458561.20128 	&	0.201 	&	96.38 	&	0.86 	&	LAMOST-MRS	&	2459325.09148 	&	0.091 	&	103.92 	&	0.91 	&	LAMOST-MRS	\\
2458561.21795 	&	0.218 	&	113.85 	&	0.92 	&	LAMOST-MRS	&	2459325.10746 	&	0.107 	&	116.49 	&	0.93 	&	LAMOST-MRS	\\
2458561.23392 	&	0.234 	&	122.92 	&	0.85 	&	LAMOST-MRS	&	2459325.12412 	&	0.124 	&	121.48 	&	0.91 	&	LAMOST-MRS	\\
2458561.25058 	&	0.251 	&	125.02 	&	0.95 	&	LAMOST-MRS	&	2459334.08116 	&	0.081 	&	-91.78 	&	0.72 	&	LAMOST-MRS	\\
2458568.14068 	&	0.141 	&	-80.07 	&	0.72 	&	LAMOST-MRS	&	2459334.09783 	&	0.098 	&	-97.23 	&	0.69 	&	LAMOST-MRS	\\
2458568.15734 	&	0.157 	&	-56.84 	&	0.70 	&	LAMOST-MRS	&	2459334.11380 	&	0.114 	&	-93.18 	&	0.69 	&	LAMOST-MRS	\\
2458568.17331 	&	0.173 	&	-26.11 	&	0.75 	&	LAMOST-MRS	&	2459337.04275 	&	0.043 	&	89.12 	&	0.91 	&	LAMOST-MRS	\\
2458568.18929 	&	0.189 	&	3.83 	&	0.76 	&	LAMOST-MRS	&	2459337.05872 	&	0.059 	&	107.03 	&	0.92 	&	LAMOST-MRS	\\
2458568.20595 	&	0.206 	&	37.06 	&	0.80 	&	LAMOST-MRS	&	2459337.07539 	&	0.075 	&	119.18 	&	0.90 	&	LAMOST-MRS	\\
2458568.22192 	&	0.222 	&	67.27 	&	0.77 	&	LAMOST-MRS	&	2459339.05372 	&	0.054 	&	-83.76 	&	0.75 	&	LAMOST-MRS	\\
2458568.23789 	&	0.238 	&	94.70 	&	0.84 	&	LAMOST-MRS	&	2459339.06969 	&	0.070 	&	-66.32 	&	0.76 	&	LAMOST-MRS	\\
2458919.20835 	&	0.208 	&	-61.83 	&	1.36 	&	LAMOST-MRS	&	2459339.08635 	&	0.086 	&	-38.33 	&	0.97 	&	LAMOST-MRS	\\
2458919.22432 	&	0.224 	&	-79.32 	&	1.10 	&	LAMOST-MRS	&	2459353.04082 	&	0.041 	&	-50.02 	&	0.87 	&	LAMOST-MRS	\\
2458919.24376 	&	0.244 	&	-97.98 	&	0.99 	&	LAMOST-MRS	&	2459353.05748 	&	0.057 	&	-68.92 	&	0.74 	&	LAMOST-MRS	\\
2458922.18330 	&	0.183 	&	53.51 	&	1.11 	&	LAMOST-MRS	&	2459353.07345 	&	0.073 	&	-86.35 	&	0.76 	&	LAMOST-MRS	\\
2458922.19928 	&	0.199 	&	79.45 	&	1.07 	&	LAMOST-MRS	&	2459952.34691 	&	0.339 	&	-72.50 	&	0.78 	&	TNO-2.4	\\
2458922.21594 	&	0.216 	&	104.43 	&	1.04 	&	LAMOST-MRS	&	2459952.40703 	&	0.510 	&	34.80 	&	0.77 	&	TNO-2.4	\\
2458943.13614 	&	0.136 	&	-60.02 	&	0.76 	&	LAMOST-MRS	&	2459952.44458 	&	0.617 	&	101.56 	&	0.87 	&	TNO-2.4	\\
2458943.15489 	&	0.155 	&	-81.43 	&	0.69 	&	LAMOST-MRS	&	2459974.37526 	&	0.945 	&	78.77 	&	0.77 	&	TNO-2.4	\\
2458943.17155 	&	0.172 	&	-92.93 	&	0.71 	&	LAMOST-MRS	&	2459974.41077 	&	0.046 	&	-39.51 	&	0.65 	&	TNO-2.4	\\
2458952.11902 	&	0.119 	&	105.12 	&	0.93 	&	LAMOST-MRS	&	2459974.46402 	&	0.198 	&	-88.66 	&	0.77 	&	TNO-2.4	\\
2458952.13499 	&	0.135 	&	119.18 	&	0.93 	&	LAMOST-MRS	&	2456701.83312 	&	0.833 	&	-77.07 	&	1.06 	&	SDSS	\\
2458952.15166 	&	0.152 	&	126.38 	&	0.83 	&	LAMOST-MRS	&	2456707.82781 	&	0.828 	&	-83.20 	&	1.06 	&	SDSS	\\
2458970.01719 	&	0.017 	&	22.07 	&	1.39 	&	LAMOST-MRS	&	2456734.74643 	&	0.746 	&	139.85 	&	1.13 	&	SDSS	\\
2458970.03316 	&	0.033 	&	59.09 	&	1.30 	&	LAMOST-MRS	&	2456757.66725 	&	0.667 	&	101.98 	&	1.30 	&	SDSS	\\
2458970.04982 	&	0.050 	&	80.78 	&	1.54 	&	LAMOST-MRS	&	2456701.83312 	&	0.833 	&	255.56 	&	6.55 	&	SDSS	\\
2458982.04267 	&	0.043 	&	115.97 	&	0.88 	&	LAMOST-MRS	&	2456707.82781 	&	0.828 	&	268.16 	&	7.48 	&	SDSS	\\
2458982.05933 	&	0.059 	&	124.17 	&	0.85 	&	LAMOST-MRS	&	2456734.74643 	&	0.746 	&	-218.23 	&	4.88 	&	SDSS	\\
2458982.07530 	&	0.075 	&	125.50 	&	0.82 	&	LAMOST-MRS	&		&		&		&		&		\\
		\hline
	\end{tabular}
    \label{table_RV}
\end{table}

\bibliography{sample631}{}
\bibliographystyle{aasjournal}



\end{document}